\documentclass[12pt,preprint]{aastex}
\usepackage{emulateapj5}

\newcommand{\etal}{et~al.\ }

\newcommand{\flu}{\hbox{erg~cm$^{-2}$}}
\newcommand{\lumin}{\hbox{erg~s$^{-1}$}}

\newcommand{\be}{\begin{equation}}
\newcommand{\ee}{\end{equation}}
\newcommand{\ba}{\begin{eqnarray}}
\newcommand{\ea}{\end{eqnarray}}

\newcommand{\batse}{BATSE}
\newcommand{\swift}{\emph{Swift}}
\newcommand{\hetetwo}{\emph{HETE-2}}
\newcommand{\sax}{\emph{BeppoSAX}}

\slugcomment{Accepted by ApJ}
\shorttitle{}
\shortauthors{}

\begin{document}

\def\Mesz{M\'esz\'aros}
\def\sarc{$^{\prime\prime}\!\!.$}
\def\arcsec{$^{\prime\prime}$}
\def\ls{\lower 2pt \hbox{$\;\scriptscriptstyle \buildrel<\over\sim\;$}}
\def\gs{\lower 2pt \hbox{$\;\scriptscriptstyle \buildrel>\over\sim\;$}}

\title{A Global Test of a Quasi-Universal Gamma-Ray Burst Jet Model
Through Monte-Carlo Simulations}

\author{Xinyu Dai\altaffilmark{1,2} and Bing Zhang\altaffilmark{3,2}}

\altaffiltext{1}{Department of Astronomy,
Ohio State University, Columbus, OH 43210, 
xinyu@astronomy.ohio-state.edu}

\altaffiltext{2}{Department of Astronomy and Astrophysics,
Pennsylvania State University, University Park, PA 16802,
xdai@astro.psu.edu, bzhang@astro.psu.edu}

\altaffiltext{3}{Department of Physics,
University of Nevada, Las Vegas, NV 89154, bzhang@physics.unlv.edu}

\begin{abstract}
The possibility that long gamma-ray burst (GRB) jets are structured
receives growing attention recently, and we have suggested that most
GRBs and their softer, less energetic fraternity, X-ray flashes
(XRFs), can be understood within a quasi-universal structured jet
picture, given that the jet structure of each individual burst is a
Gaussian or similar function. Here we perform a global test on such a
quasi-universal Gaussian-like structured jet by comparing Monte-Carlo
simulation results with a broad spectrum of observational data.  Using
the same set of input parameters as in the previous work (Zhang et
al. 2004), we confront the model with more observational constraints.
These constraints include the burst redshift distribution, jet break
angle distribution, two-dimensional redshift vs. jet break angle
distribution, luminosity function, and ${\log} N - {\log} P$
distribution.  The results indicate that the model is generally
compatible with the data. This conclusion, together with our previous
tests with the observed jet break angle vs. isotropic energy and
observed peak energy vs. fluence relations, suggests that current long
GRB and XRF data are generally consistent with such a
quasi-standard-energy and quasi-standard-angle jet picture.  With
future homogeneous burst samples (such as the one to be retrieved from
the {\em Swift} mission), the refined GRB jet structure can be further
constrained through a global comparison between various observed and
predicted burst property distributions and relations.
\end{abstract}

\keywords{gamma rays: bursts - jets}

\section{Introduction}

The geometrical configuration is an essential ingredient in
characterizing and understanding astrophysical phenomena.  There is
growing evidence that long gamma-ray bursts (GRBs) are originated from
collimated jets. This has been mainly suggested by an achromatic
steepening break observed in many GRB afterglow light-curves
\citep{rh99,ku99,ha99}. This interpretation receives indirect support
from the intriguing fact that the geometry-corrected total energy in
the GRB fireball is essentially constant (Frail \etal 2001; Bloom,
Frail, \& Kulkarni 2003), i.e., $E_{jet}=E_{iso} (1-\cos\theta_j)
\sim$ const, where $E_{iso}$ is the total energy emitted in the
gamma-ray band assuming isotropic emission, and $\theta_j$ is the jet
angle inferred from the light-curve breaks. In view of these facts,
there are two distinct approaches in constructing jet models.  One is
that different GRBs collimate the same total energy into different
angular openings, with the angular energy distribution within the jet
being constant \citep{rh99,fr01}. Another is a family of
quasi-universal structured jet models, where the structured jet has a
power-law or a Gaussian (or functions in more general forms) angular
energy distribution with respect to the jet axis (Zhang \&
\Mesz~2002a; Rossi, Lazzati, \& Rees 2002; Lloyd-Ronning, Dai, \&
Zhang 2004). In the former scenario, the jet opening angle exclusively
defines the jet break angle $\theta_j$, while in the later scenario,
in most cases $\theta_j$ is interpreted as the observer's viewing
angle. It is essential to understand whether the GRB jets are
structured, and if yes, how they are structured. This has important
implications for the fundamental questions such as the total energy
budget in the explosion, the physical origin of the collimation, and
the birth rate of the GRB progenitor.

In this paper, we define a ``structured jet'' as a jet with a certain
functional angular distribution structure of energy (and possibly
Lorentz factor as well), such as the power-law function with various
indices, the Gaussian function, or numerous other possible structures
one can think of. Since it is directly connected to the $E_{iso}
\propto \theta_j^{-2}$ (Frail et al.  2001) correlation, one
particular structured jet model, i.e., the power-law jets with index
-2 (Rossi et al. 2002; Zhang \&
\Mesz~2002a), has received broad attention. 
We call this special type of jet as the ``universal jet'' following
the convention in the literature since it has the potential to
interpret all GRB data with a universal configuration. We notice that
in some papers, ``universal jets'' and ``structured jets'' have been
used interchangeably which, to our opinion, may cause confusion to the
readers, since a certain criticism to the universal jet model may not
apply to more general structured jet models with other jet
structures. The conventional top-hat jets are called ``uniform jets''
in this paper.

The need of understanding GRB jets is boosted by the recent
identification of X-ray flashes (XRFs, Heise 2003; Kippen
\etal 2003), a fainter and softer version of GRBs, as a closely
related phenomenon with GRBs. Recent observations reveal another
intriguing empirical correlation between the cosmic rest frame GRB
spectral peak energy and the isotropic gamma-ray energy, i.e.,
$E_{peak}~\propto~(E_{iso})^{1/2}$, which was identified in the
\emph{BeppoSAX} (Amati et al. 2002) or even \emph{BATSE} 
(Lloyd, Petrosian, \& Mallozzi 2000) GRB data, and was also found to
extend to the XRF regime in the 
\emph{HETE-2} data (Lamb, Donaghy, \& Graziani 2004; Sakamoto \etal
2004). Although obvious outliers (e.g. GRB 980425) exist, the
correlation is found to be valid within the time-dependent spectra of
invidual BATSE bursts (Liang, Dai, \& Wu 2004a).  This result
strengthens the empirical law, which suggests that it is related to
some intrinsic physical processes. The relation is understandable
within the currently leading GRB models if a certain correlation
between the bulk Lorentz factor and the burst luminosity is assumed
(Zhang \& \Mesz~2002b), i.e., $\Gamma \propto L^k$ with $k$ being
different values for different models. In particular, it is consistent
with the internal shock model if the bulk Lorentz factor (and hence
the internal shock radius) is insensitive to the burst luminosity,
i.e., $k=0$. In this paper, we assume that the $E_{peak}-E_{iso}$
correlation holds for the majority of GRBs and XRFs.

In addition, the
\emph{HETE-2} sample bursts also indicate another interesting fact
that the contributions to the total number of bursts from GRBs, X-ray
rich GRBs (XRGRBs), and XRFs are approximately equal (Lamb et
al. 2004). This fact presents an important criterion to test the
validity of any jet model.

The similarities between GRBs and XRFs have stimulated studies towards
unifying the GRB and XRF phenomena through different geometrical
configurations (e.g. Lamb et al. 2004; Yamazaki, Ioka, \& Nakamura
2003; Zhang et al. 2004).  Lamb et al. (2004) pointed out that the
universal jet model is inconsistent with the GRB-XRGRB-XRF number
ratios detected by
\emph{HETE-2}, and turned to suggest a uniform jet model for all
GRBs and XRFs. If one assumes a standard energy budget for all GRBs
and XRFs, such a uniform jet model unavoidably leads to the conclusion
that GRBs have very narrow jets with typical opening angle smaller
than 1 degree. The universal jet model was also tested against various
criteria recently (e.g. Perna, Sari, \& Frail 2003; Nakar, Granot, \&
Guetta 2004; Guetta, Piran, \& Waxman 2004a), and it has been found
that such a model may violate some observational constraints. However,
a pure universal jet model corresponds to a strict $E_{iso} \propto
\theta_j^{-2}$ relation. In reality, the data indicate that this
correlation is only valid in a statistical sense (Frail et al. 2001;
Bloom et al. 2003). In the $E_{iso} -
\theta_j$ plane, the afterglow data are distributed around the
$E_{iso} \propto \theta_j^{-2}$ line with moderate scatter
(Lloyd-Ronning et al. 2004). This fact alone already suggests that GRB
jets are not ``universal''. Any jet model aiming to interpret the GRB
phenomenology is at best ``quasi-universal'', i.e., different jets may
share a more or less the same structure, but the parameters to define
the structure should have some scatter around some typical values. An
important insight is that when parameter scatter is taken into
account, the jet structure is no longer obliged to be power-law with a
-2 index. Other jet structures are also allowed (Lloyd-Ronning et
al. 2004). In particular, we (Zhang et al. 2004) recently proposed a
quasi-universal model for GRBs and XRFs. In order to successfully
reproduce the right relative numbers of GRBs, XRGRBs, and XRFs, we
suggest that the jet structure in individual bursts is Gaussian-like,
or with a similar structure. This ansat was verified with a
Monte-Carlo simulation, and the model can also reproduce the $E_{iso}
- \theta_j$ relation. In view that the narrow uniform jet model (Lamb
et al. 2004) conflicts with the standard afterglow model (Zhang et
al. 2004), and that the universal jet model encounters various
difficulties (e.g. Lamb et al. 2004; Guetta et al. 2004a), we
tentatively suggest that the quasi-standard-energy and
quasi-standard-angle Gaussian-like jet model is a more plausible one
to interpret GRB and XRF data in a unified manner.

In order to prove this suggestion, the quasi-universal jet model needs
to confront a broader spectrum of data.  Since within a structured jet
model the probability of observing the jet at angle $\theta_v$ is
proportional to $\sin(\theta_v)$, many observational properties can be
predicted once the jet structure function and the variation parameters
are given.  In this paper, besides the $E_{peak}^{obs}$ vs. fluence
relation and the $E_{iso}-\theta_j$ relation we have already tested in
Zhang et al. (2004), we consider several new constraints including the
burst redshift ($z$) distribution, jet angle ($\theta_j$)
distribution, two-dimensional $z-\theta_j$ distribution, luminosity
function, and $\log N - \log P$ distribution. Some of these criteria
have been taken individually to test some jet models (e.g. Perna \etal
2003; Lloyd-Ronning \etal 2004; Lin, Zhang, \& Li 2004; Liang, Wu, \&
Dai 2004b; Nakar \etal 2004; Guetta \etal 2004a). However, none of the
previous studies performed a global test for a particular model with
all the criteria. We believe that such a global test is essential to
constrain and to finally pin down the right GRB jet structure.  Here
we perform such a test with the quasi-universal Gaussian-like jet
model (Zhang et al. 2004). The exact GRB structure may differ from the
simple Gaussian form. We take this simple structure as the starting
point to examine how well it could reproduce the data. Due to the
``quasi-universal'' nature, analytical studies may not be adequate,
and we perform a set of Monte-Carlo simulations to access the problem.

\section{Monte-Carlo Simulations}
We perform Monte-Carlo simulations for a quasi-universal Gaussian-like
jet model.  The jet structure and the input parameters that we use are
the same as those used in \citet{zh04}, where the motivation to
introduce such a jet structure is also explained.  
Below we describe the parameters of this model in more detail.  

First, we approximate the angular energy distribution of the jet as
\be
\epsilon(\theta) = \epsilon_0 e^{-\frac{\theta^2}{2\theta_0^2}}.
\label{Gaussian}
\ee
The total energy of the jet, $E_j$, is obtained by integrating
$\epsilon(\theta)$ over the entire solid angle 
\be
E_j = 4\pi \int_0^{\pi/2}\epsilon(\theta) \sin(\theta) d\theta.
\label{jetenergy}
\ee
With a small $\theta_0$, the total jet energy is approximately $E_j \sim
2\pi\epsilon_0\theta_0^2$ \citep{zm02}.
This jet structure contains two parameters, the total energy of the
jet, $E_j$, and the characteristic jet width, $\theta_0$.  The
parameters, $E_j$ and $\theta_0$, are distributed in log-normal
distributions for the simulated bursts.  This quasi-universal approach is
required to reproduce the large scatter of the $E_{iso}$--$\theta_j$
relation of GRBs \citep{ldz04}.  In particular, these two parameters
are constrained to be around (Zhang et al. 2004)
\ba
<\log(\frac{E_j}{1~\rm erg})>~\sim~51.1 \label{Ej}\\
\sigma_{\log(\frac{E_j}{1~\rm erg})}~\sim~0.3 \\
<\log(\frac{\theta_0}{1~\rm rad})>~\sim~-1.0 \\
\sigma_{\log(\frac{\theta_0}{1~\rm rad})}~\sim~0.2~.
\ea
Although the definition $E_j$ in the Gaussian structured jet model
(Eq.~[\ref{jetenergy}]) is different from that in the
uniform jet model, our best fit typical jet energy (Eq.~[\ref{Ej}]) is
consistent with the one in the uniform jet model (Bloom et al. 2003).
We have shown that this set of input parameters can roughly reproduce the
approximately equal numbers of GRBs, XRGRBs, and XRFs, the
$E_{iso}$--$\theta_j$ relation, and the $E_{peak}^{obs}$--fluence
relation. 

Second, in a structured jet model the observing angle, $\theta_v$, is
distributed as 
\be
\frac{dN(\theta_v)}{d\theta_v} = \sin(\theta_v).
\ee
The isotropic equivalent energy, $E_{iso}$, is defined as
\be
E_{iso} = 4\pi \epsilon(\theta_v).
\ee
The jet break angle, $\theta_j$, of a Gaussian jet is (Kumar \& Granot 
2003; Zhang et al. 2004; Rossi et al. 2004)
\be
\theta_j = \left\{ \begin{array} {r@{\quad:\quad}l} \theta_0 &
\theta_v < \theta_0 \\ \theta_v & \theta_v \ge \theta_0 \end{array}
\right. 
\ee

Finally, the number of bursts per unit redshift, $N(z)$, is distributed as
\be
\frac{dN(z)}{dz} = \frac{R_{GRB}(z)}{1+z} \frac{dV(z)}{dz},
\ee
where $dV(z)/dz$ is the comoving volume per unit redshift and
$R_{GRB}$ is the GRB rate.  The comoving volume is obtained as 
\be
\frac{dV(z)}{dz} = \frac{4 \pi D_L^2 c}{(1+z)^2 H_0} [\Omega_m
(1+z)^3 + \Omega_k (1+z)^2 + \Omega_\Lambda]^{-1/2}, 
\ee
where $H_0 = 70~{\rm km~s^{-1}~Mpc^{-1}}$, $\Omega_m = 0.3$, $\Omega_k
= 0$, and $\Omega_\Lambda = 0.7$, and $D_L$ is the luminosity
distance. We assume that the GRB rate traces the star forming rate such
that 
\be
R_{GRB}(z) = \left\{ \begin{array}{l@{\quad:\quad}l} R_0 10^{0.75z} &
z < z_{peak} \\ R_0 10^{0.75z_{peak}} & z \ge z_{peak} \end{array}
\right. 
\ee
Here, we used the Rowan-Robinson star forming rate (Rowan-Robinson
1999, cf. Lin et al.\ 2004). 
Two sets of burst redshifts are simulated with the parameter
$z_{peak} = 2$ and $z_{peak} = 1$, respectively.

We simulate 10,000 bursts. For each burst we simulate $E_j$,
$\theta_0$, $\theta_v$, and $z$ according to the distributions those
parameters follow.  We calculate other parameters of the simulated
bursts in the following equations.  The peak energy of the simulated
bursts are calculated through the $E_{iso}$--$E_{peak}$ relation
(Amati et al. 2002)
\be
E_{peak} \sim 100~{\rm keV}(\frac{E_{iso}}{10^{52}~{\rm erg}})^{1/2},
\ee
in which we introduced lognormal scatter (with $\sigma \sim 0.3$) to
reflect the statistical nature of the observed data points. 
The observed peak energy is related to the rest-frame peak energy by
\be
E_{peak}^{obs} = E_{peak}/(1+z),
\ee
and the burst energy fluence is calculated as
\be
F = \frac{E_{iso} (1+z)}{4 \pi D_L^2}~.
\label{fluence}
\ee
In order to obtain the peak flux or peak luminosity of the bursts, 
we simulate the conversion time scale $T$ in the rest frame of the
bursts as a log-normal distribution (see also Lamb et al. 2004), 
\ba
<\log(\frac{T}{1~\rm s})>~\sim~0.56 \label{ct} \\
\sigma_{\log(\frac{T}{1~\rm s})}~\sim~0.05~,
\ea
such that
\ba
L_{peak} = E_{iso}/T \\
f_{peak} = \frac{F}{T(1+z)}~.
\ea
The central value of this distribution, $3.63~s$, is consistent with
the value of $3.41~s$ obtained from the \hetetwo\ and \sax\ burst
sample \citep{ldg04}.
We note that this conversion time scale should be shorter than the
true duration of the burst since the peak flux is higher than average
flux.  
The exact central value used in this paper is obtained as the best 
fit to the BATSE peak flux distribution through a series of simulations
with different central values (please see $\S$~3.5).

The simulated bursts are filtered by different detection thresholds
that represent the sensitivities of different surveys.  In principle,
the thresholds should be calculated in unit of peak photon flux.  The
peak photon flux depends on the spectral shape of the bursts, and
the sensitivity is also a function of photon energy.
Moreover, most of the bursts observed are strongly variable.  For
simplicity, here we use the simulated burst
fluence to define various detection thresholds in most of
the simulations except for the $\log N - \log P$ distribution.

\section{Results}
After obtaining the sample of the simulated bursts with various burst
parameters calculated, we compare the simulation results with current
observations and perform a global test to the GRB jet structure with
the observational constraints.  The constraints include the burst
redshift distribution, jet break angle distribution, 
redshift vs. jet break angle two-dimensional distribution, 
luminosity function, $\log N - \log P$ distribution, 
$E_{iso}$ -- $\theta_j$ relation, $E_{peak}^{obs}$ -- fluence
relation, and the relative numbers of GRBs, XRGRBs, and XRFs. We note 
that the last three tests have been presented in a previous paper
\citep{zh04}.  Here, we only present the rest of the simulation results. 

\subsection{Redshift Distribution}
There are two major uncertainties for the simulated burst redshifts.
First, we assume that the GRB rate trace the star forming rate which
may involve some uncertainties.  In
addition, the underlying star forming rate itself is not accurately
constrained.  The observed redshift distribution of the bursts can be
used, as a first step, to constrain the star forming rate used in the
simulation.  We note that the jet structure will also affect the shape
of the redshift distribution due to the limited sensitivities of the
detectors.  A homogeneous sample would be most suitable for these
studies.  The current sample of the bursts with redshift measurements
is, however, small and inhomogeneous. Nonetheless, we compare this
sample with the simulation results to obtain some preliminary
understanding of the issue. 

\begin{figure}
\plottwo{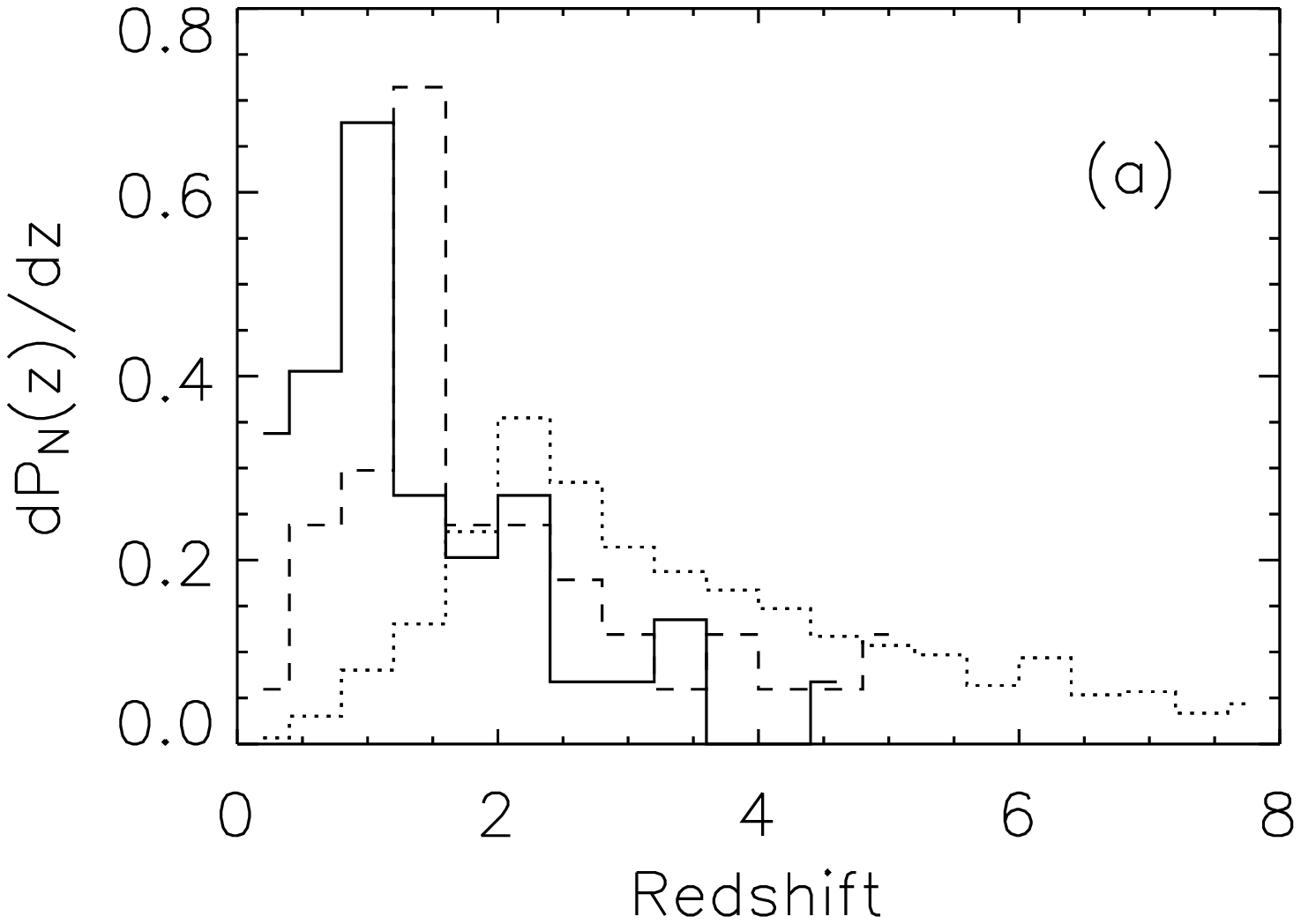}{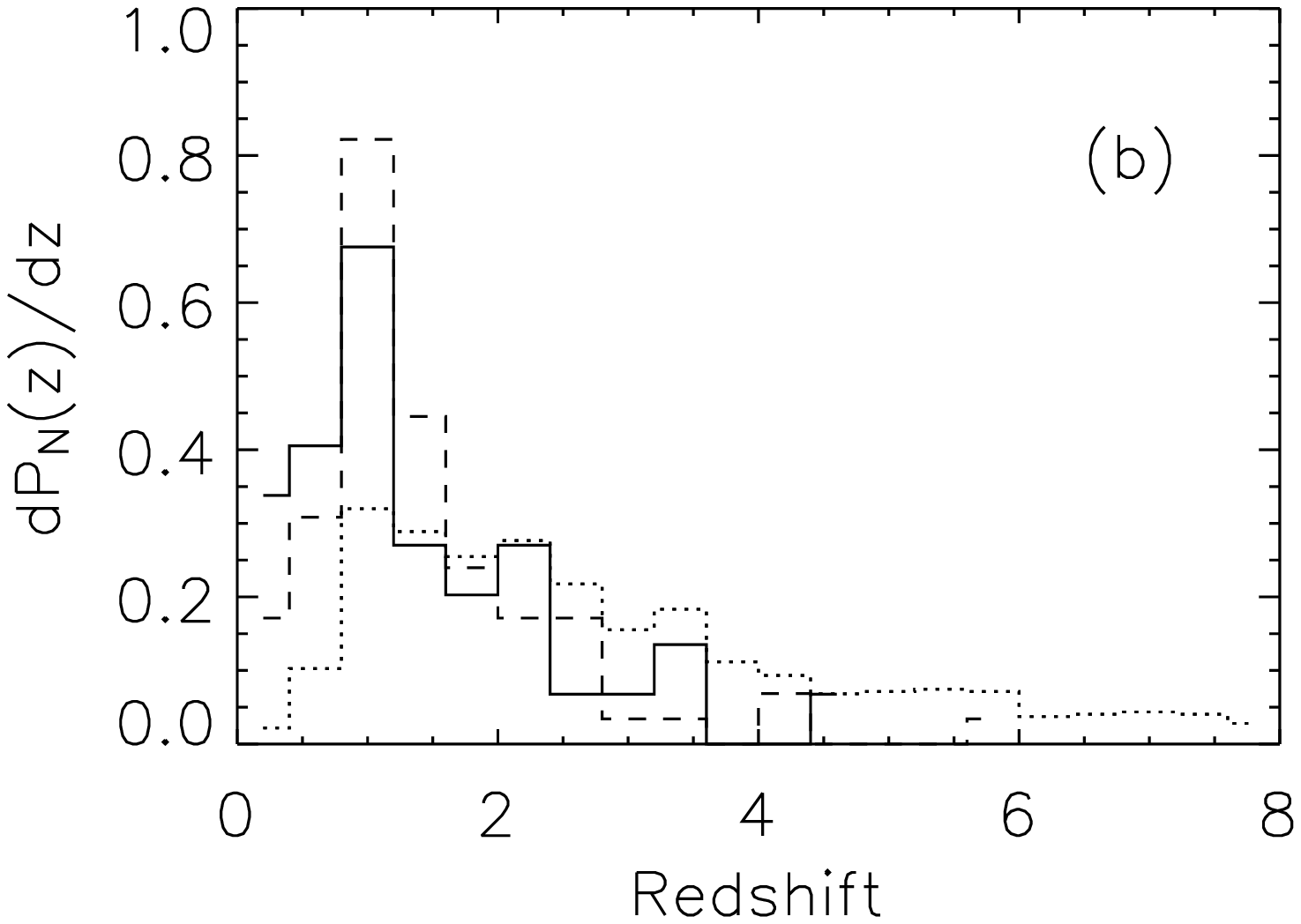}
\caption{(a) The probability distribution functions (PDF) of GRBs with
respect to redshift for the observed bursts (solid line) and the simulated
bursts with different detection thresholds for a $z_{peak}=2$
Rowan-Robinson star forming rate.  In particular, the dotted and
dashed lines are simulated distributions with fluence threshold of
5$\times$10$^{-8}$ and 1$\times$10$^{-5}$~\flu, respectively. (b) 
Same with (a), but for $z_{peak}=1$.\label{fig:dndz}}
\end{figure}

We plot the redshift distributions of the simulated bursts and compare
with the sample obtained from the observations
(Figure~\ref{fig:dndz}).  Figures~\ref{fig:dndz}a and 
\ref{fig:dndz}b represent simulations with two
different star forming rate, with $z_{peak} = 2$ and $z_{peak} = 1$,
respectively.  The observed burst redshifts are obtained from 
\citet{bfk03} and the GRB Localization website\footnote{The website is
at http://www.mpe.mpg.de/\~{}jcg/grbgen.html} maintained by
J. Greiner, and there are 37 bursts with measured redshifts in total.
We select the simulated bursts with different detecting thresholds.
The thresholds are set to be 1.0$\times$10$^{-5}$~\flu\ and
5$\times$10$^{-8}$~\flu.  The first threshold is selected to be
much higher than the triggering thresholds for most detectors in order to
reflect the selection effect that only a small number of bright GRBs have
their redshifts measured.
For example, the median fluence value of the 28 bursts with redshift
measurements in \citet{bfk03} is 2.3$\times$10$^{-5}$~\flu. 
The second threshold is selected to match the sensitivity of \emph{HETE-2}. 

\begin{figure}
\plottwo{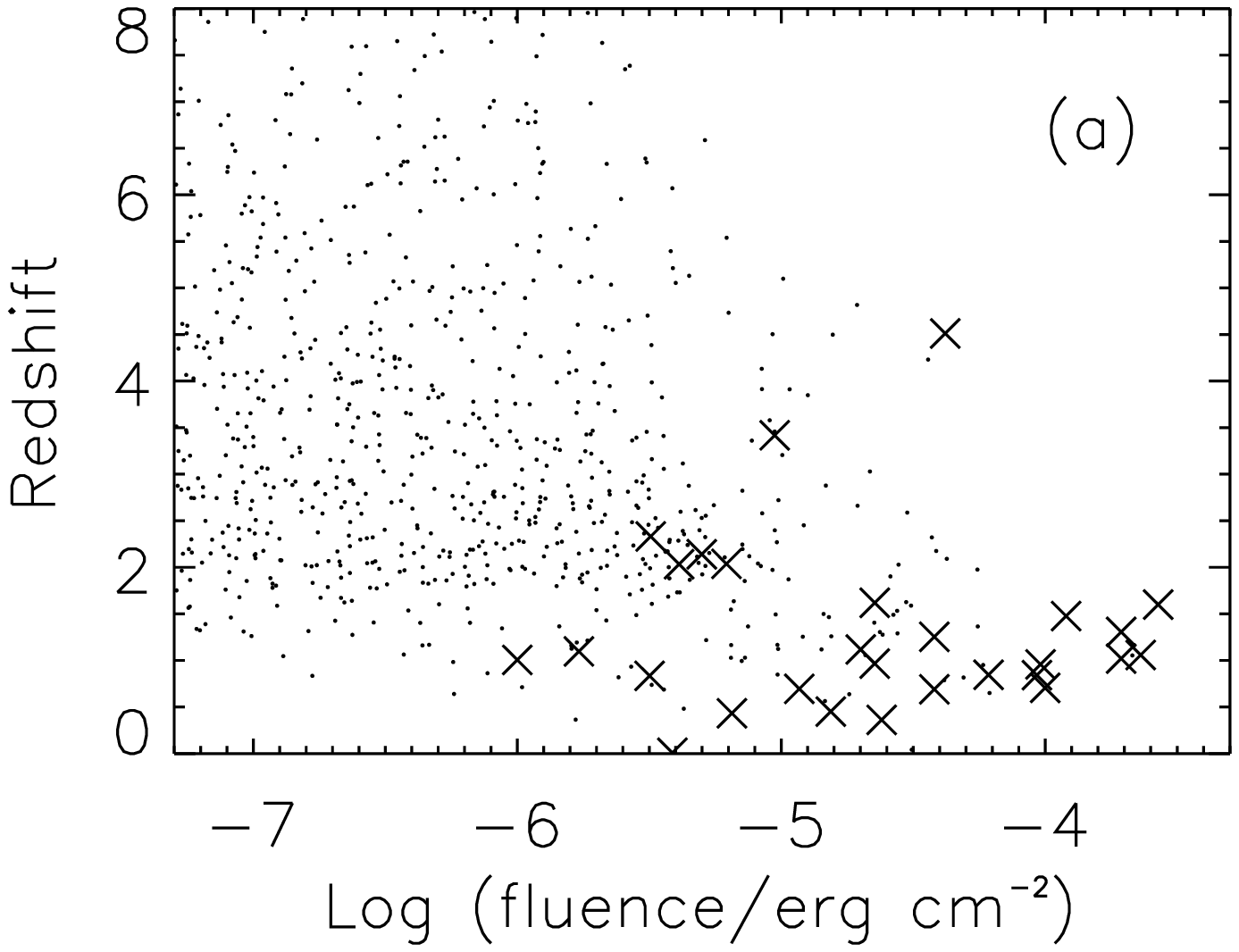}{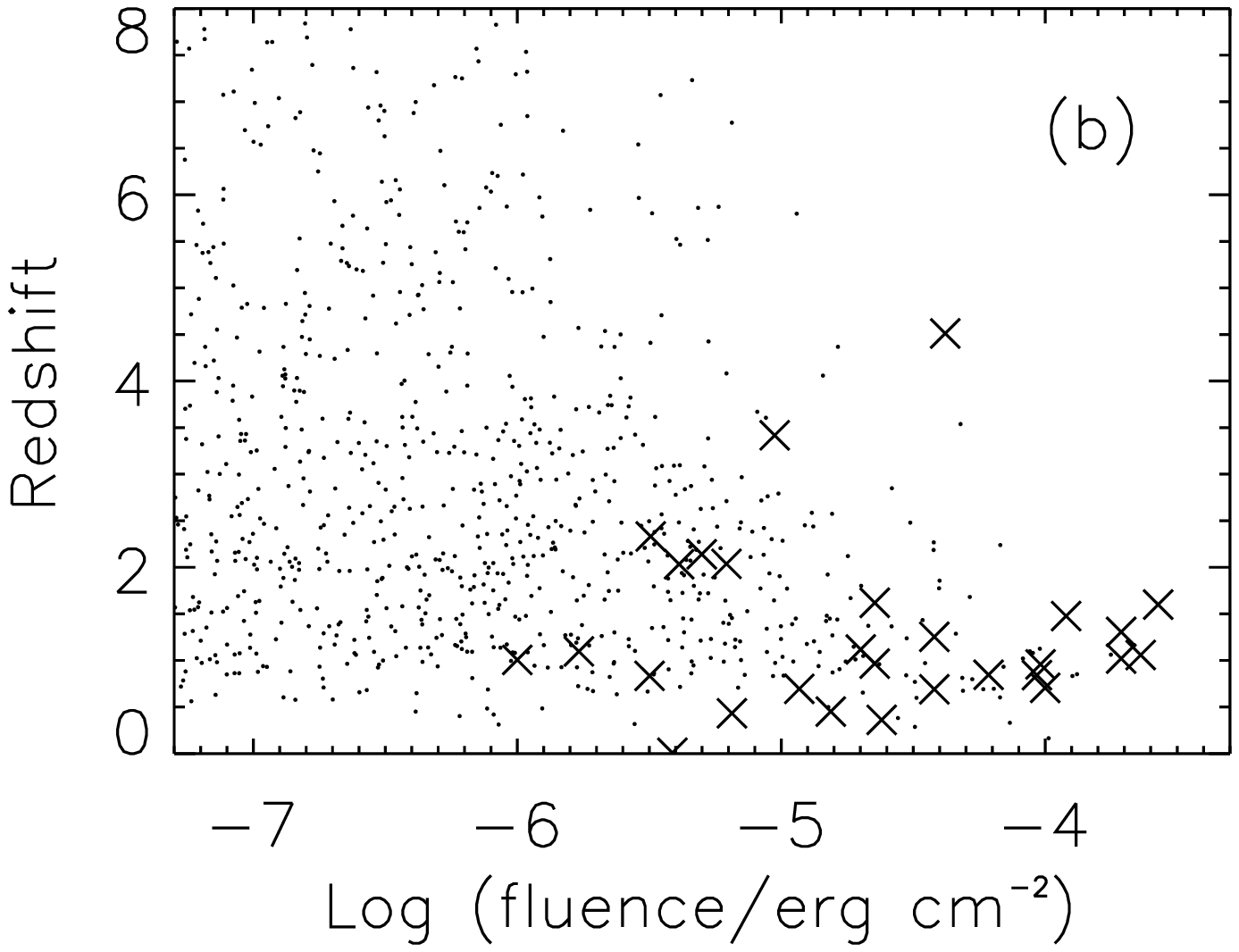}
\caption{(a) Redshift vs. fluence plot for the simulated bursts (dots)
and the observed bursts (crosses) from \citep{bfk03} with a $z_{peak}=2$
Rowan-Robinson star forming rate.
(b) Same with (a), but for $z_{peak}=1$.\label{fig:rf}}
\end{figure}

Figure~\ref{fig:dndz} shows that the observed redshift distribution
peaks at $z\sim1$.  The observed distribution is consistent with both
simulated distributions with different $z_{peak}$, if a high
detection threshold of 1.0$\times$10$^{-5}$~\flu~ is adopted. 
This high threshold is reasonable to apply here when comparing the
simulated redshift distribution with the currently observed redshift
distribution because only a few bursts have their redshifts measured. 
These bursts only account for a small fraction of the total amount of
bursts detected, and they are typically brighter.
In order to better illustrate this point, we show in
Figure~\ref{fig:rf} the redshift vs. fluence plots for the simulated
bursts and the observed bursts from \citet{bfk03} 
sample. Both star forming rate models have been plotted. 
We can see that most of the observed bursts are concentrated in the
high fluence end compared with the simulated bursts.
The current observed redshift sample cannot distinguish between the
two star forming rates. However, as a more homogeneous GRB redshift
sample is accumulated (with future instruments such as \swift), the
peak of the GRB rate can be constrained as the two models predict
different peaks as the detection threshold decreases. 
We adopt $z_{peak}=2$ when discussing the simulation results below.
An even larger and more homogeneous sample is needed in order to
constrain the shape of star forming rate, especially whether the star
forming rate is above or below the Rowan-Robinson rate at large
redshifts ($z>2$). For $z_{peak}=2$, the current data indicate
tentative evidence that the nearby events are more abundant than what
is expected from the model that assumes the standard star-forming
rate. For $z_{peak}=1$, the current model meets the data in the
low-redshift regime as well. Whether there exist extra nearby GRBs
is of great interest in the GRB community, and more redshift data are
needed before a firm conclusion is drawn.

We note that most XRFs do not have redshift measurements, which may
present a bias in the observed sample.  However, this should not
affect the result too much because the current redshift sample of GRBs
is small (37 in total).  Considering the number ratio of GRBs to XRFs
from \emph{HETE-2} data, about 12 XRFs should be added to the sample.
If the redshift distributions of GRBs and XRFs are similar, the
addition of about 12 XRFs should not change the shape of the redshift
distribution very much.  In addition, the best fits to the observed
redshift distribution are the simulated bursts with large fluences,
and according to the $E_{peak}^{obs}$--fluence relation of the
\emph{HETE-2} bursts, these high fluence bursts ($>$
1.0$\times$10$^{-5}$~\flu) should all be GRBs.  Similar arguments also
apply to the following simulations of the jet break angle distribution
and the two-dimensional jet break angle vs. redshift distribution.

\subsection{Jet Break Angle Distribution}

\begin{figure}
\plotone{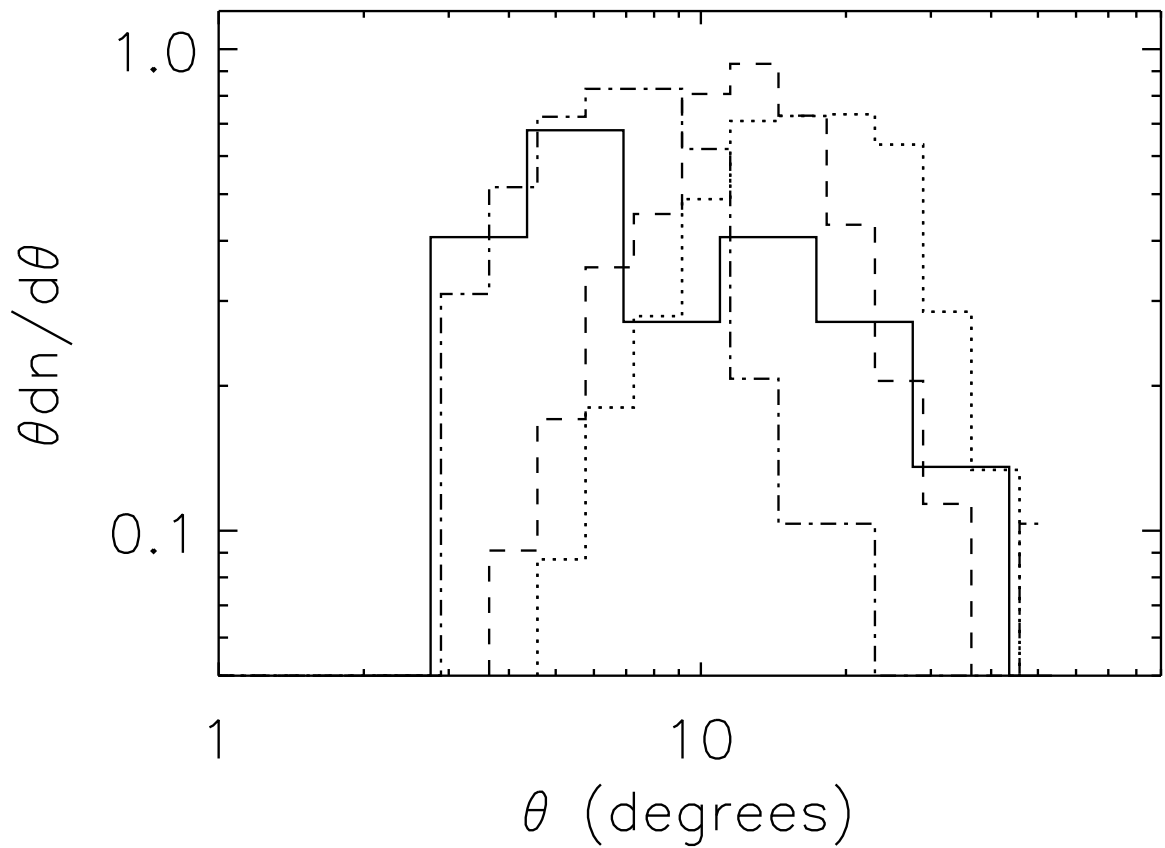}
\caption{Normalized distributions of the observed jet angles (solid
line) obtained from \citet{bfk03} and the simulated jet angles with
detection threshold of 5$\times$10$^{-8}$ (dotted line),
5$\times$10$^{-7}$ (dashed line), and 1$\times$10$^{-5}$~\flu\
(dash-dotted line), respectively.  The reason to choose a high fluence
limit, such as 1$\times$10$^{-5}$~\flu, is that only a small fraction
of total bursts have their jet break angles measured.  The median
fluence of the 16 bursts with jet break angle measurements is
2.3$\times$10$^{-5}$~\flu\ \citep{bfk03}, even higher than the highest
fluence threshold we adopt.
\label{fig:tndt}}
\end{figure}

\begin{figure}
\plotone{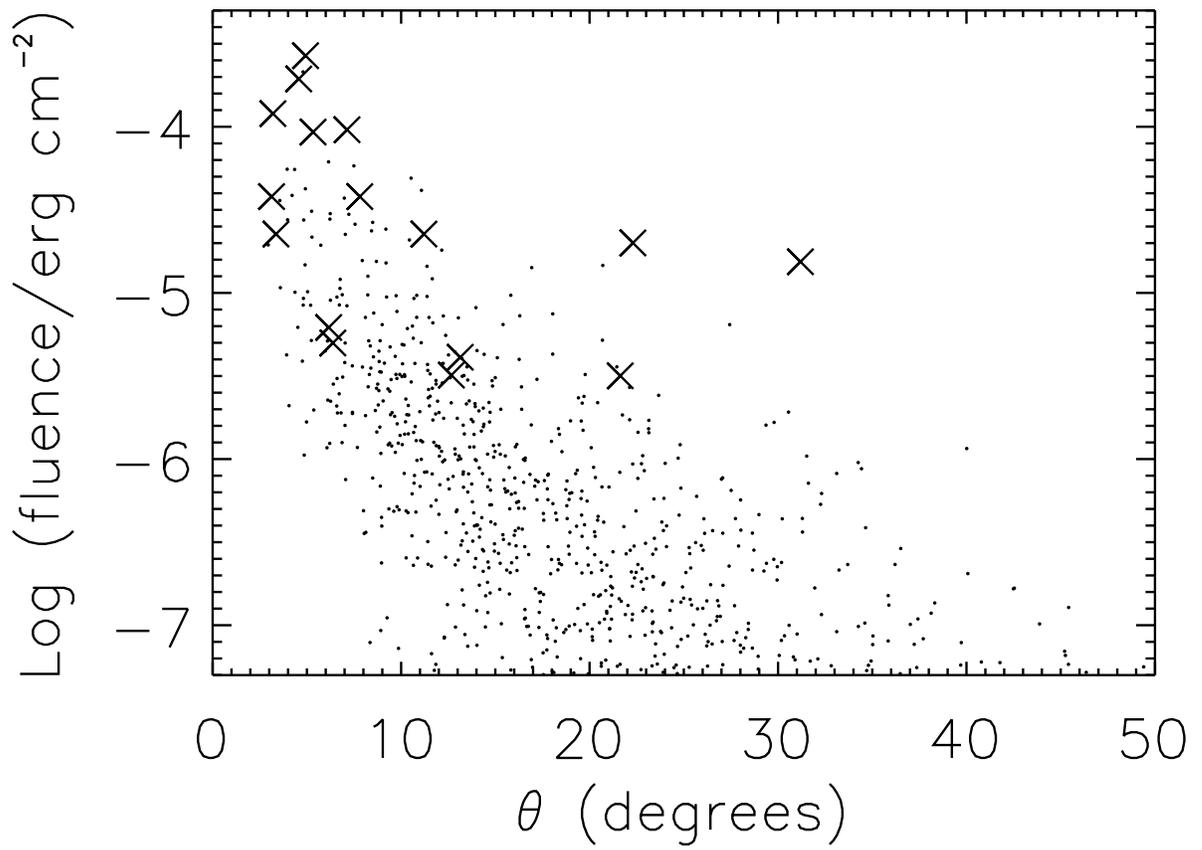}
\caption{Fluence vs. jet break angle plot for the simulated bursts
(dots) and the observed bursts (crosses) from \citep{bfk03} with a $z_{peak}=2$
Rowan-Robinson star forming rate.
\label{fig:fth}}
\end{figure}

We plot the distribution of the jet break angle of the simulated
bursts with different detection thresholds and the observed jet break
angle distribution in Figure~\ref{fig:tndt}.  The observed jet break
angles are obtained from \citet{bfk03}, and there are 16 bursts with
jet break angle measurements.  We exclude the bursts with upper or
lower limit measurements on the jet break angles when comparing with
the simulation results.  The shape of the observed jet break angle
distribution is sensitive to the bin size chosen because of the small
sample size.  However, the peak of the jet angle distribution can
roughly be constrained at about 7 degrees.  In this simulation, we
also apply different fluence filters to the simulated bursts to
simulate the effect of the limited detector sensitivities.  In
particular, we adopt the fluence thresholds of 1.0$\times$10$^{-5}$,
5.0$\times$10$^{-7}$, and 5.0$\times$10$^{-8}$~\flu~, respectively.
Figure~\ref{fig:tndt} shows that the simulated jet break angle
distribution is also sensitive to the detection threshold used in the
simulation.  The peak of the simulated distribution will move to
larger angles when a lower threshold is selected.  This is consistent
with the result of \citet{psf03}, who discovered this effect from the
universal jet model.  The simulated distribution with a fluence limit
of 1.0$\times$10$^{-5}$~\flu\ is consistent with the observed jet
break angle distribution.  Considering the difficulties in identifying
the optical afterglows and in measuring the jet break angles, such a
high threshold is a reasonable choice when comparing the simulation
with the observations.  The median fluence of the 16 bursts with jet
break angle measurements is 2.3$\times$10$^{-5}$~\flu\ \citep{bfk03},
even higher than the highest fluence limit we adopt.  Only five of the
16 bursts have fluences lower than 1.0$\times$10$^{-5}$~\flu, and the
faintest one has a fluence of 3.17$\times$10$^{-6}$~\flu.  Considering
that there are some bursts that are brighter than the fluence limit we
adopt but whose jet break angles are still not measured, it is more
reasonable to compare the median fluence, rather than the smallest
fluence, of the 16 bursts.  In Figure~\ref{fig:fth}, we show the
fluence vs. jet break angle plot for the simulated bursts and the
observed bursts \citep{bfk03}. Most of the observed bursts are in high
fluence regions.  We also compare the jet break angle distribution
with the predictions from the $z_{peak} = 1$ Rowan-Robinson star
forming rate, and lead to similar results.

We reach a similar conclusion as \citet{psf03} that the predicted jet
break angle distribution of structured jets is consistent with the
currently observed sample distribution. While Perna et al. (2003)
discussed the universal jets, our simulations are for Gaussian-like
jets.  We note that the detection threshold used in \citet{psf03} is
the 90$\%$ efficiency peak flux threshold for BATSE, which is much
more sensitive than the threshold we have adopted in the simulation
that results in consistency between the data and the simulation.
Since only a small fraction of bright GRBs have jet break angle
measurements, it may be more appropriate to use a higher threshold
than the BATSE detection threshold.  When a higher threshold is
selected, the $\theta_j$ distribution peak for the universal jet model
should move to a smaller value comparing with the observed one.
Liang et al. (2004b) also noticed this independently and also adopted a
high threshold in their simulations.

\subsection{Redshift vs. Jet Break Angle}
The jet break angle distribution discussed previously is a
one-dimensional distribution which includes bursts at all redshifts.
As pointed out by \citet{ngg04}, a more accurate test is to perform a
two-dimensional ($z-\theta_j$) distribution comparison between the
data and the model prediction.  It is possible that the
two-dimensional distribution does not agree with the observations
while by integrating over redshift the one-dimensional distribution
agrees with the observations by chance. 

\begin{figure}
\plotone{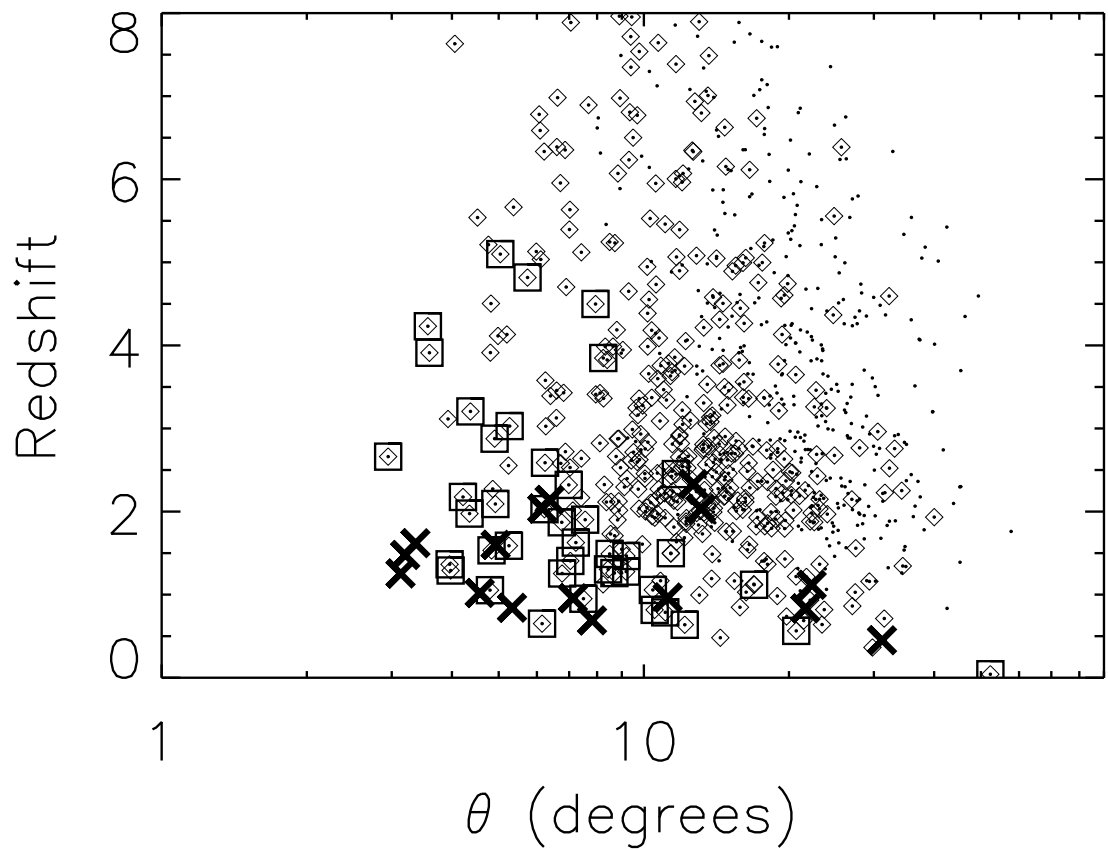}
\caption{Redshift vs. jet angle plot for the observed burst sample (crosses)
obtained from \citet{bfk03} and the simulated bursts with detection
threshold of 5$\times$10$^{-8}$ (dots), 5$\times$10$^{-7}$
(diamonds), and
1$\times$10$^{-5}$~\flu\ (squares), respectively.\label{fig:zt}}
\end{figure}

In Figure~\ref{fig:zt}, we plot the simulated data points with
different fluence thresholds (same as those used in the one
dimensional plot) together with the 16 bursts with both redshift and
jet break angle measured \citep{bfk03}.  The density of the data
points represents the probability density function (PDF) of this
two-dimensional distribution.  The plot shows that the PDF depends on
the threshold of the detector.  As the threshold goes higher, the peak
of the PDF moves towards the region containing smaller jet break
angles.  This is consistent with 
the result from the one-dimensional analysis.  In particular, the
distribution of simulated bursts with the highest threshold is consistent
with the observational distribution.  Even we limit the data points
from a narrow redshift bin ($0.8 < z < 1.7$) that includes most of
the observational data points, the simulated distribution and
observational distribution is still consistent.  Again, a high
threshold is reasonable in this analysis since the bursts with
jet break angle measurements are much brighter on average than the
ones in the whole sample of the detected bursts.

\citet{ngg04} argued that this two-dimensional distribution from the
prediction of a universal jet does not agree with the observation,
especially for the bursts within the redshift range of $0.8 < z < 1.7$.
As we have shown previously (Lloyd-Ronning et al. 2004), the
$E_{iso}-\theta_j$ data already require that the model has to be
quasi-universal, thus the inconsistency suggested by Nakar et
al. (2004) is largely due to their adopting a non-varying universal
jet model. In particular, the sharp boundary of the region allowed by
the universal model should smooth out if a quasi-universal model is
considered (as shown in Figure~\ref{fig:zt}). Moreover, the
detection threshold used in \citet{ngg04} is also the BASTE detection
threshold, which is much more sensitive compared with the burst sample
with jet break angle measurements. Also we adopted a Gaussian-like jet
structure, while they stick to the power-law structure which suffers
other problems (e.g. numbers of XRFs respect to GRBs and the $\log N -
\log P$ distribution) as well.

\subsection{Luminosity Function}
\begin{figure}
\plotone{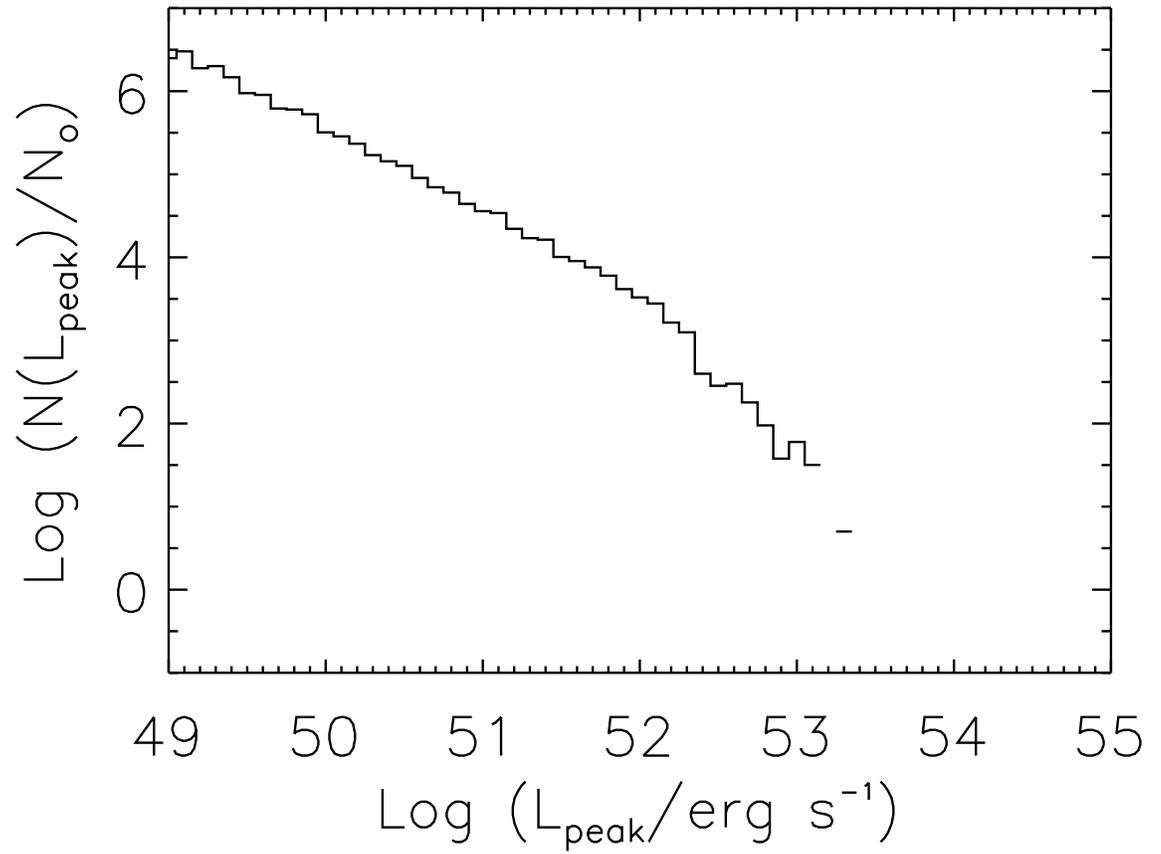}
\caption{The luminosity function of GRBs with a quasi-universal
Gaussian-like jet structure.  The simulated luminosity function can be
characterized by a broken power law, with the power law indices of
$\sim -2$ in the high luminosity part ($L > 10^{52}~\lumin$) and $\sim
-1$ in the low
luminosity part ($L < 10^{52}~\lumin$). This is consistent with that
obtained from \citet{sc01}, except for the low luminosity 
end below $10^{50}~\lumin$.\label{fig:lum}}
\end{figure}

We plot the luminosity function of the present quasi-universal
Gaussian-like jet model in Figure~\ref{fig:lum}.  This luminosity
function is an update of the result presented in \citet{ldz04}.  In
\citet{ldz04}, the total energy of the jet was taken as a constant
rather than quasi-universal. When the total energy scatter is
introduced, the simulated luminosity function does not show
a bump at the break of the powerlaw index (cf. Figures 9 and 10 in
Lloyd-Ronning et al. 2004).  In addition, we have performed
a rigorous calculation of the total energy instead of using the
approximation for small angles. At large angles, this difference is
more than a factor of two.

The simulated luminosity function can be characterized by a broken
power law, with a power law index of $\sim -2$ at high luminosity end
($L > 10^{52}~\lumin$) and an index of $\sim -1$ for low luminosities
($L < 10^{52}~\lumin$).  The simulated luminosity function is
consistent with previous simulations of Gaussian jets performed in
\citet{ldz04}, except that there is no big bump at the break of the
power index this time. Currently, the GRB luminosity function is not
directly determined from the observations, since the sample of bursts
with redshift measurement is too small. Nonetheless, there are several
attempts to constrain the luminosity function through various
approaches (e.g. Schmidt 2001; Norris \etal 2002; Lloyd-Ronning,
Fryer, \& Ramirez-Ruiz 2002; Stern, Tikhomirova, \& Svensson 2002;
Firmani et al. 2004).  In general, many of these studies found a break
of the luminosity function, with the power index steeper in the high
luminosity range and flatter in the low luminosity range (see detailed
discussion in Lloyd-Ronning \etal 2004 and references therein).  This
is consistent with the simulated luminosity function.  In particular,
the shape of the simulated luminosity function in this paper is very
similar with that obtained from \citet{sc01}, except for the low
luminosity end below $10^{50}~\lumin$ where the luminosity functiom
from \citet{sc01} turns over.  This luminosity function obtained from
\citet{sc01} can fit the BASTE $\log N - \log P$ quite well assuming a
certain star forming rate \citep{sc03}.

\subsection{$\log N - \log P$ Distribution}
We use the $\log N - \log P$ plot as a final test to the jet structure
used in this paper.  The $\log N - \log P$ plot has been used to
constrain the jet opening angle distribution for the uniform jet and
the star forming rate in previous studies \citep{lzl04, gpw04}.  In
particular, \citet{gpw04} pointed out that the $\log N - \log P$
distribution predicted by the 
universal jet model over predicts bursts with faint flux. This
inconsistency is another manifestation of the problem of
over-producing XRFs in the universal jet model (Lamb et
al. 2004). Since the quasi-universal Gaussian jet model can overcome
the latter difficulty, it is natural to expect that it can solve the
former problem as well.

We use the bursts from the offline re-analyzed BATSE catalog \citep{ko00}
including both triggered and untriggered bursts.  The catalog includes
a total of 2167 bursts, in which 1393 are triggered bursts and 874
are untriggered bursts.  
The bursts selected in this catalog are all long GRBs, which are directly
related to our model.
In order to estimate the simulated peak photon flux, we assume a Band
function \citep{ba93} for the simulated bursts, and adopt the low- and
high-energy photon indices as 
$\alpha = -1$ and $\beta=-2$. 
As our simulation stems from the data provided by \citet{bfk03} where
the isotropic energy is given in the 20--2000~keV band rest frame, we
need to take into account the difference between the bandpass of BATSE
and that used in \citet{bfk03}. 
The fluence that we obtained from Eq.(\ref{fluence}) should be
\be
F = \int_{\rm 20/(1+z) keV}^{\rm 2000/(1+z) keV} E N(E) dE~.
\ee
The photon fluence in the BATSE band (50--300 keV) reads
\be
F_{\rm BATSE}^{ph} = \int_{\rm 50 keV}^{\rm 300 keV} N(E) dE~. \\
\ee 
We can then calculate the photon fluence in the BATSE band 
using the energy fluence, i.e. 
\be
F_{\rm BATSE}^{ph} = \frac{F \int_{\rm 50 keV}^{\rm 300 keV}
N(E) dE} {\int_{\rm 20/(1+z) keV}^{\rm 2000/(1+z) keV} E N(E) dE}~. 
\label{Fph}
\ee
The ratio between the two
integrals in Eq.(\ref{Fph}) only depends on the value of $E_p$ (which
is simulated in the code) when both $\alpha$ and $\beta$ are assigned
to their typical values\footnote{We have also tested the cases when $\alpha$
and $\beta$ deviate from the nomical values. The results essentially
remain unchanged.}. 
Essentially, we use the energy fluence to determine the normalization of the
Band function and calculate the photon fluence with the determined Band function.
Finally, the peak photon flux can be obtained by
dividing $F_{\rm BATSE}^{ph}$ by the conversion timescale $T$ defined in
Eq.(\ref{ct}), i.e.,
\be
f_{\rm BATSE}^{ph} = \frac{F_{\rm BATSE}^{ph}}{T(1+z)}~.
\ee

\begin{figure}
\plotone{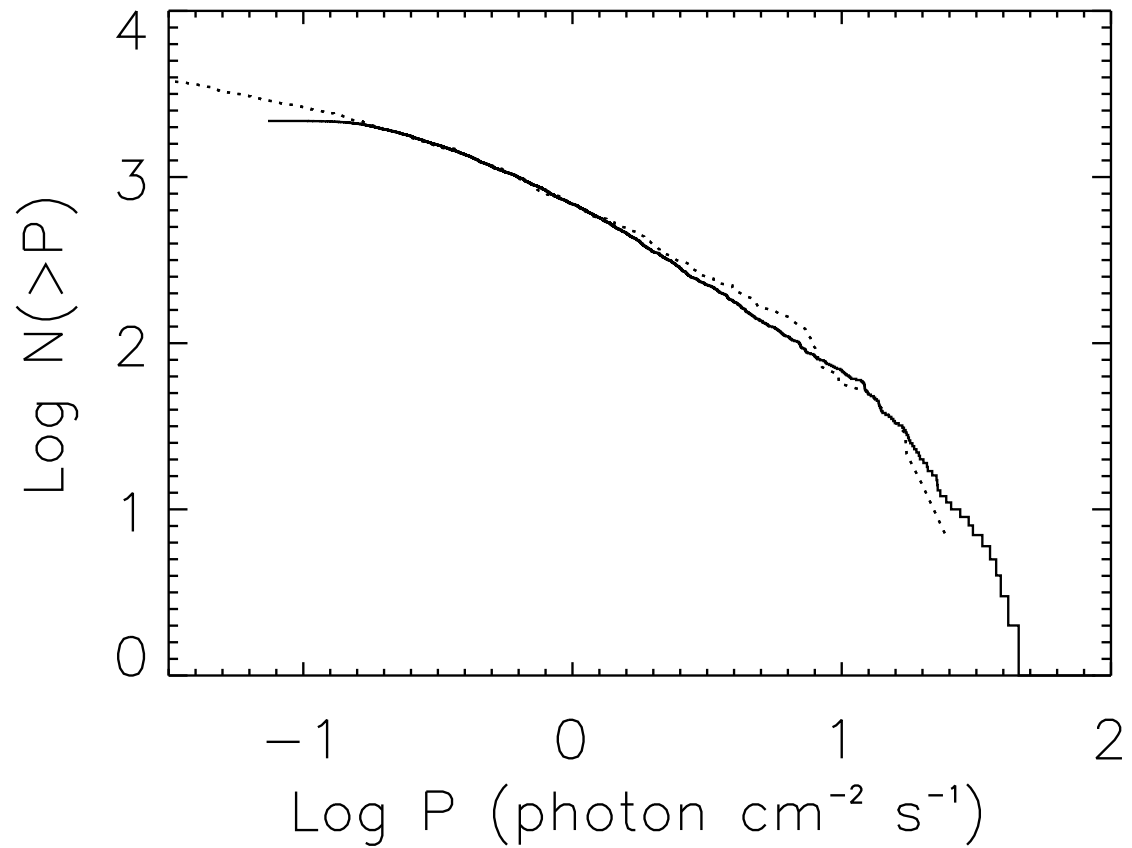}
\caption{$\log N(>P)- \log P$ plots for the BATSE bursts (solid histogram, 
from Kommers \etal 2000) and the simulated bursts from the
quasi-universal Gaussian like model (dotted line).
The univeral jet fails to reproduce the $\log N(>P)- \log P$
distribution \citep{gpw04} as it predicts almost a power law $\log
N(>P)- \log P$ distribution that unavoidably over-predicts bursts at
the low flux end. The simulation results from the quasi-univeral
Gaussian jet fits well with the BATSE $\log N(>P)- \log P$
distribution.
\label{fig:lnlp}}
\end{figure}

We do not perform a fluence or energy truncation for the $\log N -
\log P$ analysis (unlike in the previous sections). 
The BATSE sample is much more homogeneous, and is free of
most of the selection effects encountered for the other samples we have
discussed (the sample with redshift or jet angle information). The
simulated sample is natually truncated with the offline detection threshold
in peak photon flux \citep{ko00}, 
 i.e., $0.18~{\rm photon~cm^{-2}~s^{-1}}$. This is lower
than the BATSE onboard detection threshold, $0.3~{\rm
photon~cm^{-2}~s^{-1}}$. 
This peak photon flux truncation is adopted when we compare the simulated
$\log N - \log P$ distribution with the observation through the
Kolmogorov-Smirnov (K-S) test. Guided by the K-S test,
we fit the simulated $\log N - \log P$ distribution to the observed
distribution with a normalization parameter so that the 
simulated $\log N - \log P$ plot can be shifted vertically.
The normalization reflects the difference between the number of the
simulated burst and the number of the true bursts in the BATSE sample.
After the fitting, the simulated bursts fit very well with the
observed $\log N - \log P$ plot. The $\log N - \log P$ distributions
for the simulated bursts after the fitting and the bursts from BATSE
catalog are shown in Figure~\ref{fig:lnlp}.  We perform a K-S test to
the two cumulative distributions after the peak photon flux truncation
 at $0.18~{\rm photon~cm^{-2}~s^{-1}}$ and a K-S chance probability of
11$\%$ is obtained, which implicates that the simulated $\log N - \log
P$ distribution cannot be rejected through the K-S test. 

We also test the model with the $z_{peak}=1$
Rowan-Robinson star forming rate and similar results are obtained.
This result suggests that the observed BATSE $\log N - \log P$
distribution can be reproduced with this quasi-universal Gaussian
structured jet model. We also compare the BATSE catalog from
\citet{ko00} with the GUSBAD catalog from \citet{sc04}, and the $\log
N - \log P$ distributions obtained from the two catalog are generally
consistent.

The simulated $\log N - \log P$ plot can generally reproduce the
turnover behavior at the faint end, but still deviates the BATSE data
at the lowest peak photon flux of the distribution.  This does not
affect the main result because this deviation occurs at regions where
the peak photon flux is below the limit of the BATSE offline search
sensitivity so that a significant part of the burst could be missed.
In addition, at low flux region the distribution is sensitive to the
detection threshold adopted.  We plot the simulated distribution
extending down to a lower peak photon flux limit than BATSE as a
prediction for future observations.  In the simulation we did not
perform any luminosity or $E_{peak}$ truncation so that XRFs are also
included in the simulated samples for this test.  It is unclear if
most XRFs have been detected with BATSE.  Since a Band function
\citep{ba93} have a typical upper photon spectral index $\beta \sim
-2$ for GRBs, XRFs should be detected as faint GRBs as well if their
high energy photon spectral indices have a similar value.  In
addition, we also test the case by excluding the simulated burst with
lower observed peak energies (e.g. those with $E_{peak}^{obs}<10$
keV), and the resulting $\log N - \log P$ plot is still consistent
with the data. The reason is that XRFs mainly contribute to the
faintest population of the distribution.

We note that the simulated $\log N - \log P$ plot is also sightly
different from the observed distribution at high photon fluxes.
However, the difference is not significant because the number
difference between the simulated and observed bursts is small.
The difference could result from small number statistical
effects. This could be tested by increasing the number of simulated
bursts. It may also indicate that the jet structure is more close to a
power-law structure at small angles.

\citet{gpw04} fitted the predicted $\log N - \log P$ distribution from
the universal jet model and compared it with the observations.  The
total bursts (595) used in \citet{gpw04} is smaller than the burst
sample (2167 in total) used in this paper.  However, even with a small
sample, \citet{gpw04} concluded that the universal jet model can be
rejected from the $\log N - \log P$ distribution.  With a
quasi-universal Gaussian-like jet model, the observed $\log N - \log
P$ distribution can be fitted quite well from the simulations
performed in this paper, especially in the range of $ -0.7 <\log
P~({\rm photon~cm^{-2}~s^{-1}})< 1$ where the $\log N - \log P$
distribution is constrained most accurately.  The major difference
between a power-law jet and a Gaussian jet is at large angles. The
exponential drop of the jet energy at large angles in the Gaussian
model is the key to reduce the right number of faint bursts in the
BATSE sample.  The problem faced by the universal jet model in the
$\log N - \log P$ distribution test is avoided with the
quasi-universal Gaussian jet model.

Recently \citet{ggb04} claims that a power law jet truncated at large
angles can also reproduce the BATSE $\log N - \log P$ distribution.
This is consistent with our argument raised in this paper. In view
that a Gaussian jet invokes a natural exponential drop off at large
angles while a reasonable power-law jet model has to invoke both a
small angle and a large angle artificial break, we deem that the
quasi-universal Gaussian jet model is a more elegant one. We notice
that the $\log N - \log P$ distribution is more sensitive to the jet
structure than to the GRB distribution with redshift. It is therefore
a powerful tool to pin down the correct jet structure.

\section{Conclusion and Discussion}

We perform Monte-Carlo simulations of a quasi-universal Gaussian-like
structured jet and compare the simulation results with a wide spectrum 
of current observations.  The simulation results for the Gaussian-like
jet used in the paper are generally consistent with various
observational constraints, including the burst redshift distribution,
jet break angle distribution, two-dimensional distribution of redshift
and jet break angle, luminosity function, and $\log N - \log P$
distribution. This result is complementary to the previous simulation
results \citep{zh04} that showed that the number ratios among GRBs,
XRGRBs and XRFs, the observed jet break angle vs. isotropic
energy relation, and the observed peak energy vs. fluence relation are
consistent with predictions from this jet model.

Although the samples of different burst properties used for some tests
in this paper are small so that very detailed constraints on the jet
structure cannot be achieved, the global test performed on the
quasi-universal Gaussian jet model (including this paper and the
previous work, Zhang et al. 2004) at least suggests the following two major
conclusions.  First, in order to unify GRBs and XRFs through viewing
angle effects such that GRBs are viewed from small observing angles
and XRFs are viewed from large observing angles, the jet energy at
large angles must drop exponentially.  This comes from two
constraints, i.e., the $\log N - \log P$ distribution and the number
ratios among GRBs, XRGRBs and XRFs.  This requirement is consistent
with the Gaussian jet model. The power-law jet structure cannot extend
to large angles.  Otherwise, it will over produce the number of XRFs
and low flux bursts in the $\log N - \log P$ distribution.  Second,
``quasi-universal'' should also be an essential ingredient for the jet
models, as it is unrealistic to assume that all the GRB progenitors,
their environments, and other properties are exactly the same.  The
quasi-universal nature is crucial to fit the $E_{iso}$--$\theta_j$
relation, and it could produce the luminosity function power-law index
break that was implicated in many studies.  A quasi-universal
Gaussian-like jet is suitable for both constraints.  We also note that
many burst property distributions are sensitive to the detection
threshold selected, and a suitable threshold for the current sample
should be selected carefully when comparing the model predictions with
the observations.

Currently, the detailed jet structure at small angles is not well
constrained.  The slope may be steeper than that in the Gaussian
model.  It is possible that the Gaussian jet slightly under-predicts
bright bursts as indicated from the $E_{peak}^{obs}$--fluence plot and
the $E_{iso}-\theta_j$ plot in \citet{zh04}. Such a deficit, however,
could be well due to selection effects since the brightest bursts are
those most easily to detect and to localize. If the deficit is real,
it may be accounted for through possible evolution of the GRB
luminosity function, which was suggested by recent studies
\citep{lfr02,wg03,yo04,gr04}. It may also be understood in terms of
the two-component jet picture \citep{be03,hu04,ld04}, where a bright core
component contribute to more bright bursts. These issues can be tested
thoroughly with a larger, homogeneous sample of bursts accumulated.

In the \swift\ era, the sample of GRBs with redshift and jet break
angle measurements is anticipated to be much larger than
the current sample. In addition, these bursts will form a homogeneous
sample that is most suitable to apply statistical analyses on various
burst properties.  We anticipate most of the GRB relations used in this
paper and \citet{zh04} will be constrained more accurately, except for 
the $\log N - \log P$ distribution. With refined Monte-Carlo
simulations of this and other models, the details of
GRB jet structure can be pinned down more precisely through statistical
analyses of the observed burst properties and the model predictions. 

\acknowledgements We thank Nicole Lloyd-Ronning, Peter \Mesz~ for
stimulative collaborations, Don Lamb, Rosalba Perna, Dale Frail, 
S. N. Zhang, for
extensive discussion on GRB jets, Maarten Schmidt, E. W. Liang,
Z. G. Dai, Dafne Guetta for helpful comments, and the anonymous
referee for helpful suggestions. We also acknowledge Jefferson
M. Kommers for providing the \batse\ catalog of triggered and
untriggered burst data.  This work is supported by Swift GI program
(Cycle 1) (for both authors), a NASA grant NAS8-01128 (for X.D.), and
a NASA LTSA program NNG04GD51G (for B.Z.).

\clearpage

\clearpage

\clearpage

\clearpage

\clearpage

\clearpage

\clearpage

\clearpage

\end{document}